\documentclass{PoS}

\usepackage{bbm,amsmath} 

\newcommand{\fm}{\, {\rm fm}}

\newcommand{\gev}{\, {\rm GeV}}

\newcommand{\MSbar}{\overline{\rm MS}}

\newcommand{\be}{\begin{equation}}
\newcommand{\ee}{\end{equation}}
\newcommand{\bea}{\begin{eqnarray}}
\newcommand{\eea}{\end{eqnarray}}

\newcommand{\bi}{\begin{itemize}}
\newcommand{\ei}{\end{itemize}}

\boldmath
\title{B-physics from lattice QCD...with a twist}

\unboldmath

\ShortTitle{B-physics from lattice QCD}
\author{
\vspace*{-5mm}
\begin{flushright}
IFIC/12-80 \\
FTUAM-12-113 \\
IFT-UAM/CSIC-12-111 \\
RM3-TH/12-20\\
LTH966\\
CERN-PH-TH/2012-328

\end{flushright}
N. Carrasco$^{(a)}$, P. Dimopoulos$^{(b,c)}$, R. Frezzotti$^{(b,c)}$, V. Gim\'enez$^{(a)}$, 
G. Herdoiza$^{(d)}$, \newline 
V. Lubicz$^{(e,f)}$, G. Martinelli$^{(g,h)}$, C. Michael$^{(i)}$, 
D. Palao$^{(c)}$, G.C. Rossi$^{(b,c)}$,\newline 
F. Sanfilippo$^{(j)}$, \speaker{A.~Shindler}\thanks{Heisenberg Fellow}$~~^{(k)}$, 
S. Simula$^{(f)}$, C. Tarantino$^{(e,f)}$
\\
$^{(a)}$Departament de F\'isica Te\`orica and IFIC, Univ. de Val\`encia-CSIC 
\vspace{0.2cm}
\\
$^{(b)}$Dipartimento di Fisica, Universit\`a 
di Roma ``Tor Vergata'' 
\vspace{0.2cm}
\\
$^{(c)}$INFN, Sezione di ``Tor Vergata'' c/o Dipartimento di Fisica, 
Universita` di Roma ``Tor Vergata'' 
\vspace{0.2cm}
\\
$^{(d)}$Departamento de F\'isica Te\`orica and Instituto de F\'isica Te\`orica UAM/CSIC, 
\vspace{0.2cm}
\\
$^{(e)}$Dipartimento di Fisica, Universit\`a Roma Tre 
\vspace{0.2cm}
\\
$^{(f)}$INFN, Sezione di Roma Tre c/o Dipartimento di Fisica, Universit\`a Roma Tre 
\vspace{0.2cm}
\\
$^{(g)}$SISSA 
\vspace{0.2cm}
\\
$^{(h)}$INFN, Sezione di Roma, 
\vspace{0.2cm}
\\
$^{(i)}$Theoretical Physics Division, Dept. of Mathematical Sciences, 
University of Liverpool
\vspace{0.2cm}
\\
$^{(j)}$Laboratoire de Physique Th\'eorique (Bat. 210), Universit\'e Paris Sud, 
\vspace{0.2cm}
\\
$^{(k)}$ CERN, Theory Group, Geneva
}

\abstract{We present a precise lattice QCD determination of the $b$-quark mass, of the $B$ and $B_s$ decay constants
and first results for the $B$-meson bag parameters. 
For our computation we employ the so-called ratio method and 
our results benefit from the use of improved interpolating operators for the $B$-mesons.
QCD calculations are performed with $N_f$ = 2 dynamical light-quarks 
at four values of the lattice spacing and the results are extrapolated to the continuum limit.
The preliminary results are
$\overline{m}_b(\overline{m}_b) = 4.35(12) {\rm~GeV}$ for the $\overline{\rm MS}$ $b$-quark mass, $f_{B_s} = 234(6) 
{\rm ~MeV}$ and $f_{B} = 197(10) {\rm ~MeV}$ for the $B$-meson decay constants,
$B_{B_s}^{\overline MS}(\overline{m}_b) = 0.90(5)$ and $B_{B}^{\overline MS}(\overline{m}_b) = 0.87(5)$ for the 
$B$-meson bag parameters.
}
\FullConference{36th International Conference on High Energy Physics,\\
                July 4-11, 2012\\
                Melbourne, Australia}

\begin{document}
\section{Introduction}
\vspace{-0.4cm}
The search for New Physics (NP) together with stringent tests of the Standard Model (SM) 
pass through a detailed study of physical processes involving the $b$-quark. 
Two particularly important cases for the detection of potentially large NP contributions are the purely 
leptonic decays $B~\to~\tau~\nu_\tau$ and $B_s \to \mu^- \mu^+$.\footnote{Very recently the first measurement
of this decay rate~\cite{Aaij:2012ct} shows a good agreement with the SM prediction.} 
The relevant entries in the SM prediction for these decay rates are the CKM matrix element $V_{ub}$, 
and the pseudoscalar decay constants $f_B$ and $f_{B_s}$. 
Additionally the B-Physics parameters, 
$\Delta m_d/\Delta m_s$, $\Delta m_d$, play a crucial role in the Unitarity Triangle Analysis, 
and their determination rely on lattice QCD computations of the bag-parameters $B_{B_s}$ and $B_{B_s}/B_{B_d}$.

Despite the present (B-factories, LHCb) and future (SuperB factories) experimental programs,
lattice QCD results of hadronic parameters need to have reduced uncertainties at the level of $\sim 1 \%$
(see refs~\cite{Tarantino:2012mq,Zanotti:2012ic} for recent reviews).

In this proceeding we report on the ongoing project, within the ETMC collaboration, to compute
B-physics hadronic parameters with Wilson twisted mass (Wtm) lattice fermions.
We extend and improve our previous 
analysis~\cite{Dimopoulos:2011gx,Blossier:2009hg} in two ways:
\begin{itemize}
\vspace{-0.2cm}
\item we optimize the interpolating operators for heavy-light systems to better project onto the
fundamental state
\vspace{-0.2cm}
\item we extend the range of heavy masses, $\mu_h$, considered reaching values of approximately 
$\mu_h \sim 2.5\,m_c$, where $m_c$ is the charm quark mass.
\end{itemize}
\section{Computational details}
\label{sec:impr_op}
\vspace{-0.4cm}
We use, for this analysis, the $N_f=2$ dynamical gauge configurations with up and down mass degenerate 
quarks, $m_{u/d}=\mu_l$, generated by the European 
Twisted Mass Collaboration (ETMC)~\cite{Baron:2009wt}. 
The lattice action is the tree-level improved Symanzik gauge action~\cite{Weisz:1982zw} 
and the twisted mass quark action~\cite{Frezzotti:2000nk} at maximal twist~\cite{Frezzotti:2003ni}.
The strange and the charm quarks are quenched in this work. \newline
We have used four lattice spacings~\cite{Blossier:2010cr} 
and we take the values of the light and strange renormalized
quark masses together with the pseudoscalar density renormalization constants
$Z_P^{\MSbar}(2\gev)$ respectively from~\cite{Blossier:2010cr} and~\cite{Constantinou:2010gr}.
All the lattice details of our computation can be found in our recent lattice 2012 contribution~\cite{Carrasco:2012de}.
As usual heavy-light meson masses and decay constants are extracted studying the Euclidean time dependence
of the 2-point functions with heavy-light interpolating operators.
To improve the projection onto the fundamental state and keep the noise-to-signal ratio under control 
we extract meson masses at relatively small temporal separations.
Thus the lowest lying energy eigenstate must have a large overlap with our interpolating fields.
We have constructed these fields using the so-called Gaussian smearing~\cite{Gusken:1989qx} (see~\cite{Carrasco:2012de} for
details).
The improvement with respect with the standard local interpolating operators 
can be seen in the left plot of fig.~\ref{fig:eff_masses} where we show the Euclidean time dependence of the 
effective masses for a particular simulation
point obtained from local-local (LL), and two improved interpolating operators.
The plot describes well the general properties for heavy-light effective masses: the improved correlators
allow a safe extraction of the ground state mass at shorter Euclidean time separations compared to the 
LL correlators. 
\section{The $b$-quark mass and decay constants $f_B$ and $f_{B_s}$}
\label{sec:bmass}
\begin{figure}[!tb]
\vspace{-0.9cm}
\includegraphics[width=0.5\textwidth,angle=0]{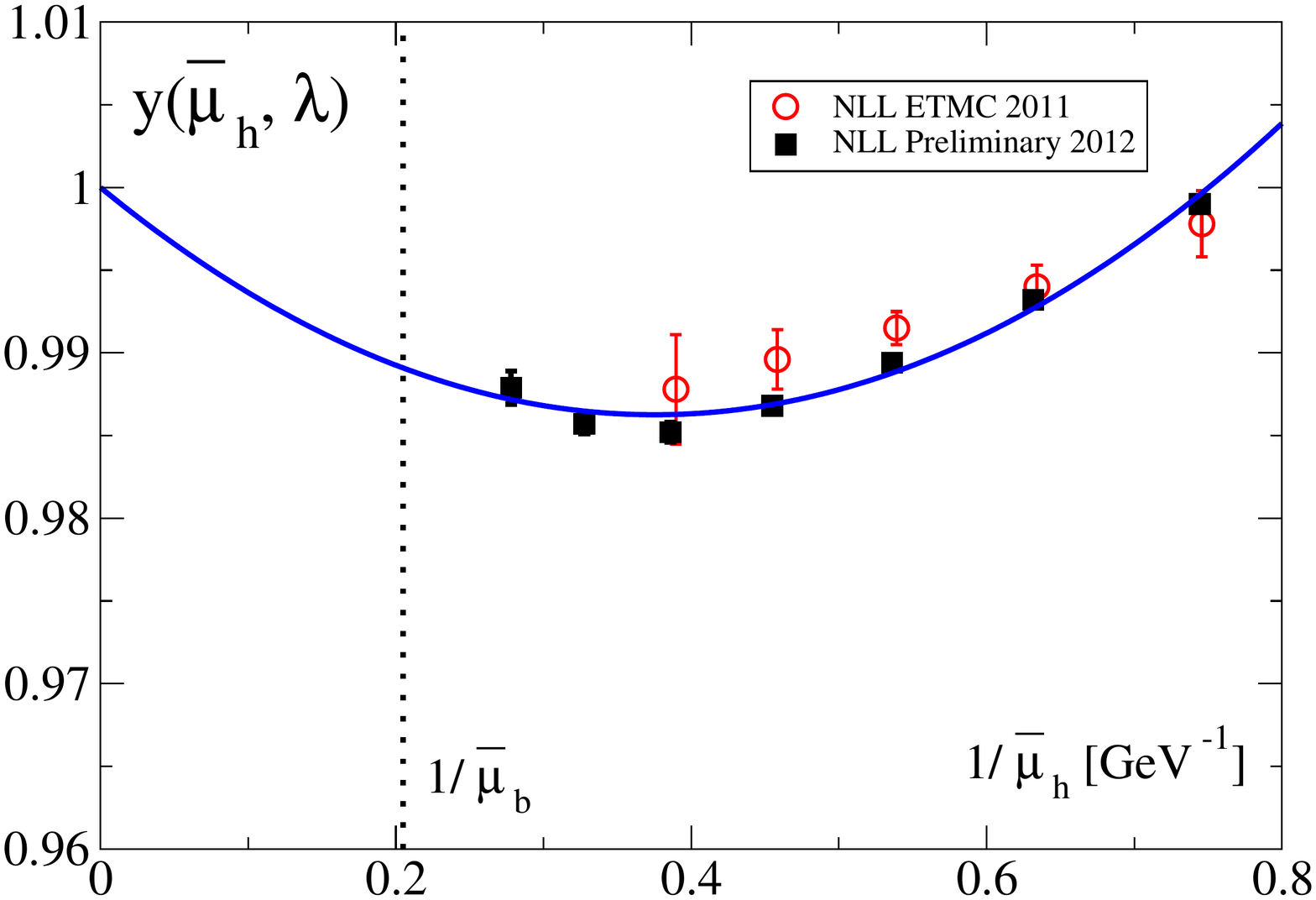}
\vspace{-0.5cm}
\includegraphics[width=0.5\textwidth,angle=0]{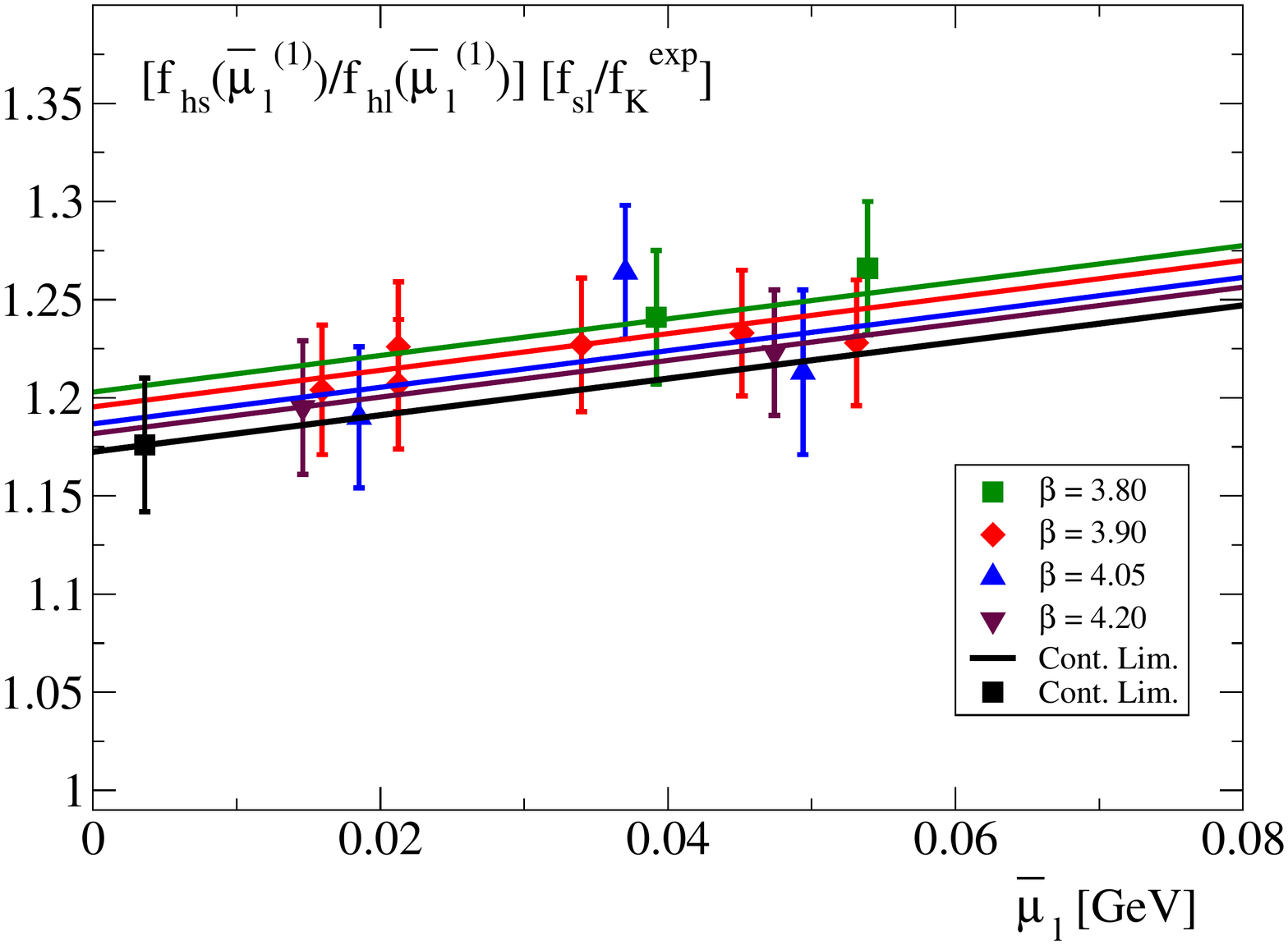}
\caption{Left plot: the continuum ratio-function $y(\overline\mu_h,\lambda)$, for $\lambda= 1.1784$ 
as a function of the heavy-quark mass.
For comparison we show our previous results~\cite{Dimopoulos:2011gx} without the interpolating operators 
improvement. Right plot: chiral extrapolation for the double-ratio 
$\left[f_{hs}/f_{hl}\right]\cdot\left[f_{sl}/f_K^{\rm exp} \right]$ 
at the triggering point, $\overline\mu_h^{(1)}$.}
\label{fig:b_mass}
\end{figure}

\vspace{-0.4cm}
To determine the B-physics quantities we implement the ratio-method 
(see ref.~\cite{Blossier:2009hg} for the details of the method).
We shortly recall here the basic idea for the computation of the $b$-quark mass. 
The same strategy applies for the other quantities computed in these proceedings. 
The static limit of the heavy-light meson mass $M_{hl}$ in the pole heavy-quark mass $\mu_h^{\rm pole}$
is given by 
\begin{figure}[tb]
\vspace{-0.9cm}
\hspace{0.3cm}\includegraphics[width=0.5\textwidth,angle=0]{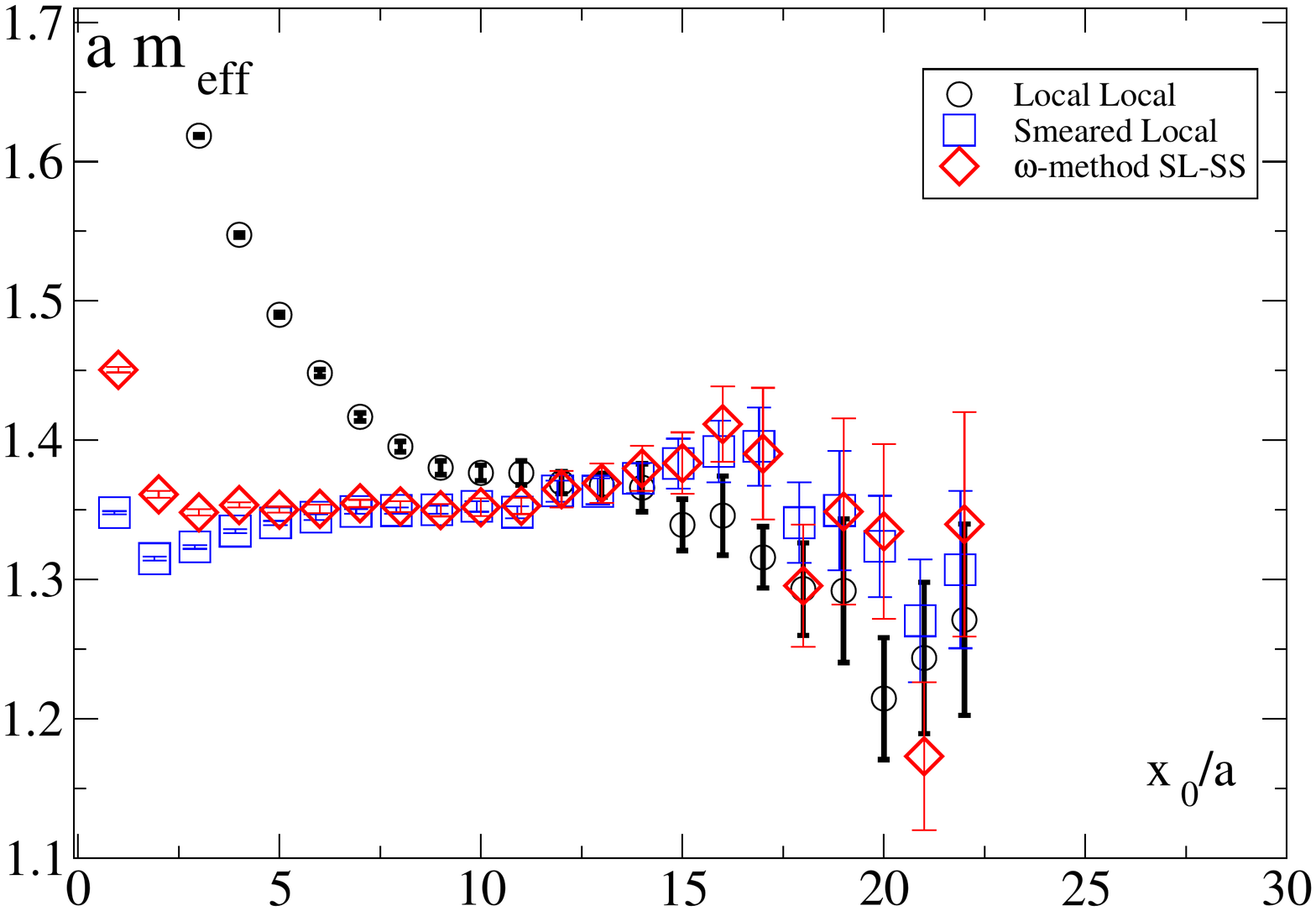}
\vspace{-0.5cm}
\hspace{0.3cm}\includegraphics[width=0.5\textwidth,angle=0]{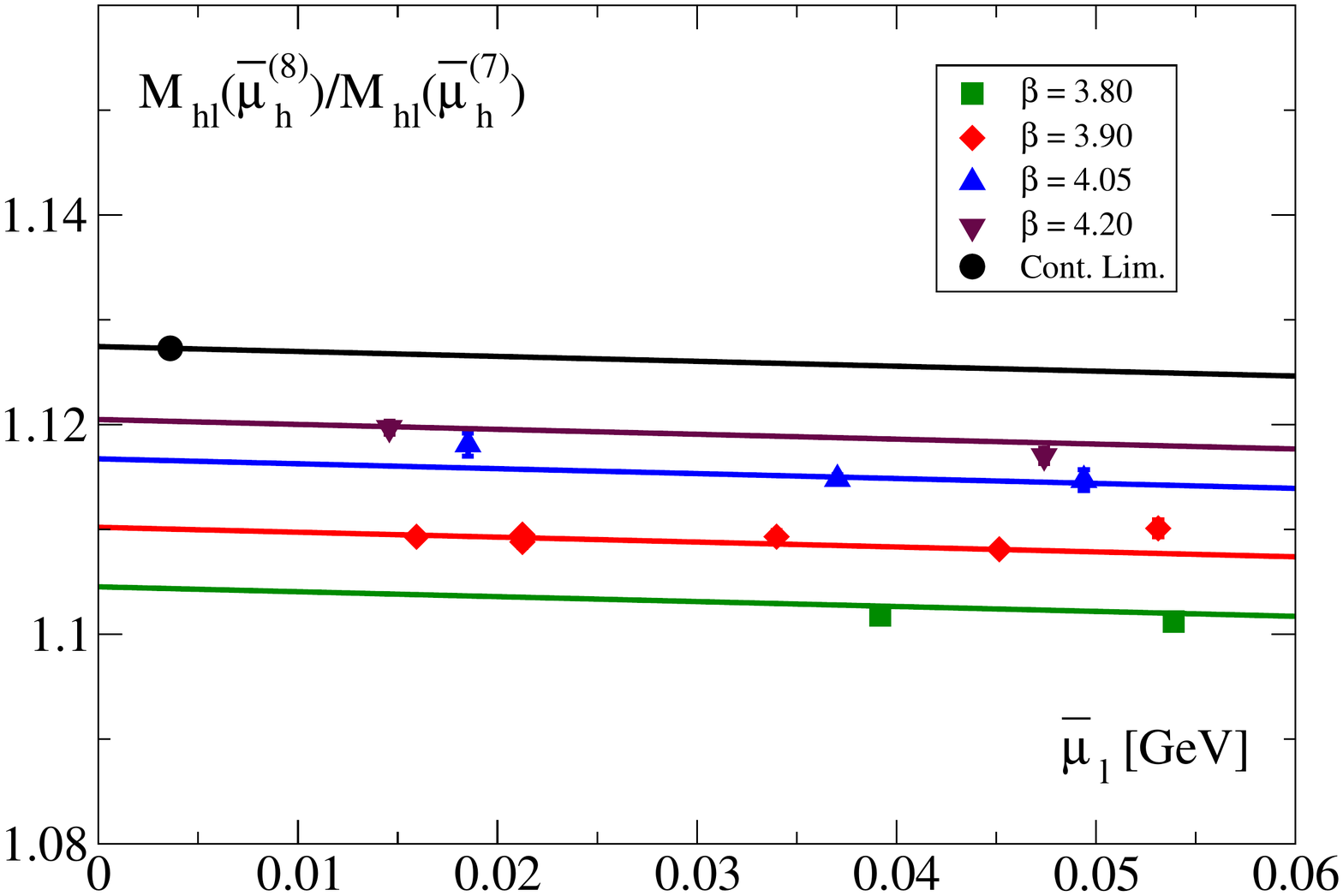}
\caption{Left plot: Euclidean time dependence of the 
effective masses at a lattice spacing of $a=
0.098(3)~\fm$, of a heavy-light correlation function
obtained from LL and improved correlation functions. The heavy-quark mass corresponds to $\sim 2.3 m_c$. 
Right plot: chiral-continuum extrapolation of the ratio 
of heavy-light meson masses at 
the heaviest quark masses, $\overline\mu_h^{(7,8)}$.}
\label{fig:eff_masses}
\end{figure}
\be
\lim_{\mu_h^{\rm pole} \rightarrow \infty}\left(M_{hl}/\mu_h^{\rm pole}\right) = 1 \,.
\label{eq:Mhl}
\ee
We take now a set of heavy-quark masses $\overline\mu_h^{(1)}<\overline\mu_h^{(2)} < \cdots<\overline\mu_h^{(N)}$
with ratio $\lambda$, i.e. $\overline\mu_h^{(n)} = \lambda \overline\mu_h^{(n-1)}$, fixed.
We denote the masses renormalized in the $\MSbar$ at a $2$ GeV scale with a ``bar''.
The ratios of heavy-light meson masses at subsequent values of the heavy-quark mass, 
$y(\overline\mu_h^{(n)},\lambda;\overline\mu_l,a)$, properly normalized, are the key 
quantities of the ratio-method~\cite{Blossier:2009hg}.
From QCD asymptotic freedom and eq.~\eqref{eq:Mhl} it follows that the ratio $y$ in the continuum limit
\be
\lim_{\overline\mu_h \rightarrow \infty} \lim_{a\rightarrow 0}y(\overline\mu_h^{(n)},\lambda;\overline\mu_l,a) = 1\,,
\ee
has an exact static limit.
The chiral and continuum extrapolation of the ratio of the heavy-light meson masses evaluated at the heaviest 
quark mass is shown in the right plot of fig.~\ref{fig:eff_masses}.
The lattice results are well described by a fit linear in $a^2$ and in the light quark mass.
We also note that discretization errors are sufficiently well under control.
As it can be seen from the left plot in fig.~\ref{fig:b_mass}, 
the ratio $y(\overline\mu_h,\lambda)$, extrapolated to the chiral 
and continuum limit, has a non-perturbative heavy-quark mass dependence that is well described by 
the HQET-inspired function
\be
y(\overline\mu_h,\lambda) = 1+\frac{\eta_1(\lambda)}{\overline\mu_h} + \frac{\eta_2(\lambda)}{\overline\mu_h^2}\,.
\label{eq:fit_ansatz}
\ee
This plot clearly shows that the accurate description of the non-perturbative dependence of the ratio $y$ 
on $\overline\mu_h$ is possible only because of the improvements that we have implemented in this analysis.
In particular the results are well described by the fit ansatz in eq.~\eqref{eq:fit_ansatz},
that provides, with the fit parameters $\eta_1$ and $\eta_2$,  
the non-perturbative description of the heavy-quark mass dependence of the ratio $y(\overline\mu_h,\lambda)$
over the whole range of heavy-masses.
The $b$-quark mass can be computed simply taking the triggering heavy-light meson mass 
$M_{hl}(\overline\mu_h^{(1)})$ around the charm region and through a repeated application 
of the ratio $y(\overline\mu_h,\lambda)$ reaching the experimental value of the $B$ meson mass $M_B$.
This is always possible with a slight tuning of $\lambda$ and $\overline\mu_h^{(1)}$ 
that leads after a 4-loop evolution to\footnote{Note that a $b$-quark mass value smaller by $\sim
120$~MeV would be found by setting $N_f=4$ (rather than $2$) in the
evolution from $2$ GeV to the $b$-scale.}
\be
m_b^{\overline{MS}}(m_b)|_{N_f=2} = 4.35(12) {\rm~GeV}\,.
\label{eq:mub}
\ee
We obtain a perfectly consistent result if we use as input the heavy-strange meson mass $M_{hs}$. 
The result we present here~\eqref{eq:mub} is in agreement with our previous determination~\cite{Dimopoulos:2011gx} with
slightly reduced error.
The $2.7\%$ relative error is dominated by the ones associated to the lattice spacing
and $Z_P$ determinations.
The statistical and systematic uncertainties coming from the application of the ratio method
turn out to be negligible, after the improvements just discussed.

A completely analogous strategy can be adopted to compute heavy-light and heavy-strange decay constants. 
We find advantageous to define ratios with exactly known static limit for the heavy-strange decay constant,
$z_s(\overline\mu_h^{(n)},\lambda;\overline\mu_l,\overline\mu_s,a)$, and for the ratio $f_{B_s}/f_B$, 
$\zeta(\overline\mu_h^{(n)},\lambda;\overline\mu_l,\overline\mu_s,a)$~\cite{Dimopoulos:2011gx}.
Both $f_{hs}(\overline\mu_h^{(1)})$ at the triggering point 
and the ratio $z_s$ have a smooth chiral-continuum extrapolation
with cutoff effects always well under control. The heavy-quark mass dependence in the continuum
of $z_s$ can again be described by a formula as the one in eq.~\eqref{eq:fit_ansatz}.
We use the $b$-quark mass, $\overline{m}_b$, we have determined 
as input leading us, using the ratio-method, to
\be
f_{B_s} = 234(6) {\rm ~MeV}\,.
\ee

The double-ratio $\zeta(\overline\mu_h^{(n)},\lambda;\overline\mu_l,\overline\mu_s,a)$ has a very smooth and weak
dependence on the light-quark mass, the lattice spacing and the heavy-quark mass.
The chiral behaviour of $f_{hs}/f_{hl}$ at the triggering mass, $\overline{\mu}_h^{(1)}$,
suggests that the heavy-quark still behaves as a relativistic one.
Thus for the chiral-continuum extrapolation
of $f_{hs}/f_{hl}(\overline{\mu}_h^{(1)})$ we find convenient to study the double-ratio
$\left[ f_{hs} / f_{hl} \right] \cdot \left[f_{sl} / f_K^{\rm exp}\right]$, for which a smooth
dependence on the light quark mass is expected~\cite{Roessl:1999iu,Allton:2008pn}.
The results are well described by a linear fit in the light-quark mass as it can be seen from the
right plot in fig.~\ref{fig:b_mass}.
This analysis of $\zeta(\overline\mu_h^{(n)},\lambda;\overline\mu_l,\overline\mu_s)$ and 
$\left[ f_{hs} / f_{hl} \right] \cdot \left[f_{sl} / f_K^{\rm exp}\right]$ leads to the values
\be
\frac{f_{B_s}}{f_B} = 1.19(5), \qquad f_B = 197(10) {\rm~MeV}\,,
\ee
where we add in quadrature the statistical and systematic uncertainties.
\section{Bag parameters}
\label{sec:bag}
\vspace{-0.4cm}

\begin{figure}[tb]
\vspace{-0.8cm}
\hspace{0.3cm}\includegraphics[width=0.5\textwidth,angle=0]{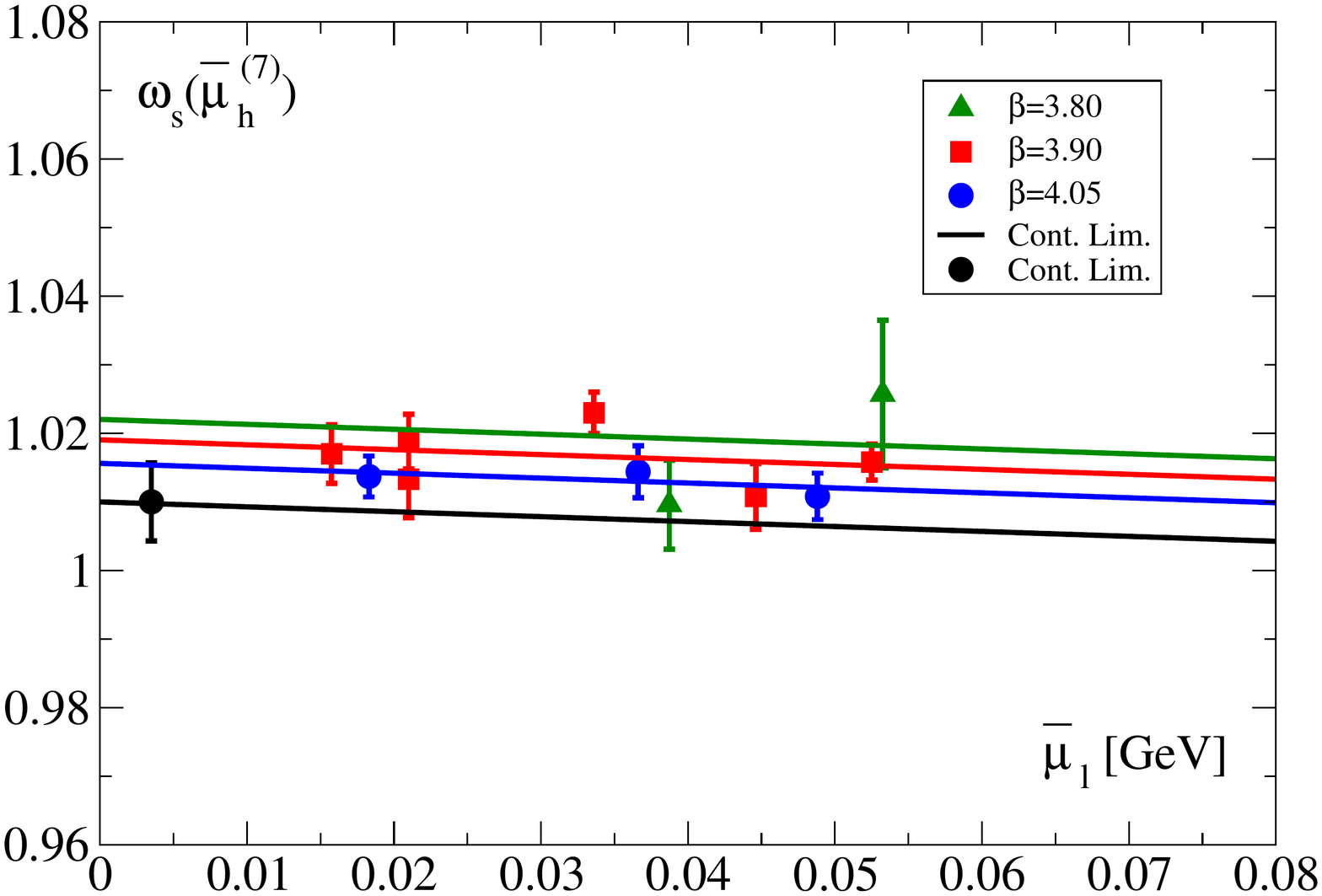}
\vspace{-0.5cm}
\includegraphics[width=0.5\textwidth,angle=0]{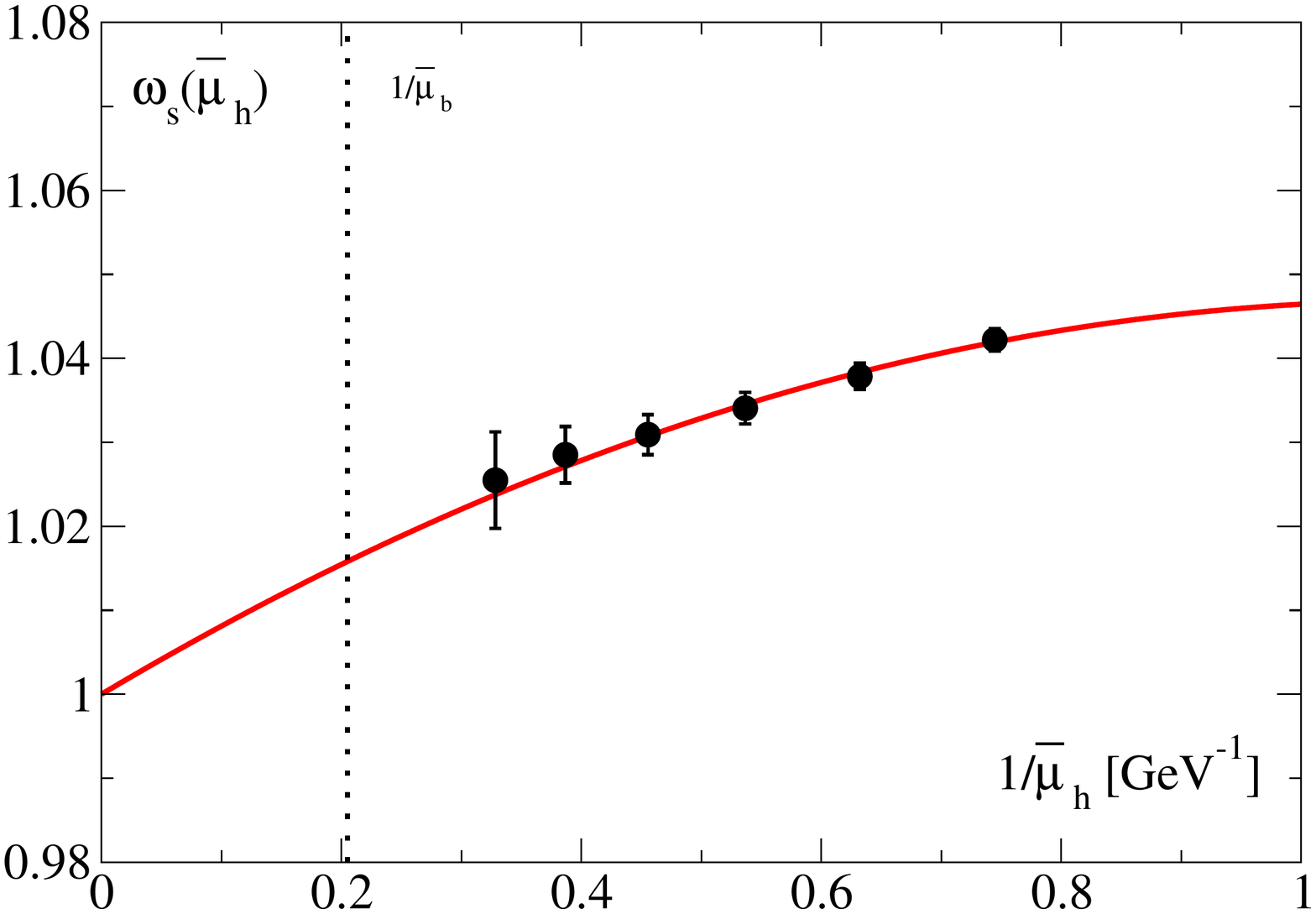}
\caption{Left plot: chiral-continuum extrapolation of the ratio 
of the bag parameters, $\omega_s$, for
the heaviest quark mass analyzed, $\overline\mu_h^{(7)}$. Right plot:
heavy-quark mass dependence of the ratio function $\omega_s(\overline\mu_h,\lambda)$, for $\lambda= 1.1784$ in the continuum and its non-perturbative description.}
\label{fig:bag}
\end{figure}

The renormalized operator $O^{\Delta B=2}_q \equiv \left[\overline{b}\gamma_\mu (1 - \gamma_5) q\right]\left[\overline{b}\gamma_\mu (1 - \gamma_5) q \right]$ in QCD is parametrized in terms of the bag parameter $B_{B_q}(\mu)$ as follows
\be
\langle \overline{B}_q | O^{\Delta B=2}_q | B_q \rangle^{\overline{MS}} \equiv \frac{8}{3}f_{B_q}^2 B_{B_q}^2(\mu) M_{B_q}^2
\ee
where $\mu$ is the renormalization scale.
With Wtm fermions one can define an operator with the correct continuum limit whose renormalization 
is multiplicative~\cite{Frezzotti:2004wz}.
Using heavy-quark symmetry one can show that the ratio of renormalized $B$-parameters evaluated in QCD 
is expected to approach unity as $1/\overline\mu_h \to 0$, with corrections of order 
$1/{\rm log}(\overline\mu_h/\Lambda_{QCD})$ for small values of $1/\overline\mu_h$.
To analyze the impact of such corrections we consider the ratio
\be
\omega_{q}(\overline\mu_h,\lambda;\overline\mu_l, a) =   
\frac{B_{B_{q}}(\overline\mu_h,\overline\mu_l, a;\mu)}{B_{B_{q}}(\overline\mu_h/\lambda,\overline\mu_l, a;\mu)}
\cdot \frac{C(\overline\mu_h;\mu^*,\mu)}{C(\overline\mu_h/\lambda;\mu^*,\mu)}\,,
\ee
where the $C$-factors ratio contains the info on the
$1/{\rm log}(\overline\mu_h)$-corrections at a fixed order in renormalization group (RG) improved perturbation theory (PT). 
Matching HQET to QCD in PT one can evaluate the function $C$.
We consider here HQET-to-QCD matching only at tree level and LL order in PT
(thereby avoiding the complications of operator mixing in HQET~\cite{Becirevic:2004ny}), and
confirm the impact of $1/{\rm log}(\overline\mu_h)$-corrections on the final results
to be at the level of one standard deviation~\cite{Carrasco:2012dd}. 
The chiral and continuum limit of the ratio $\omega_s$ at the heaviest mass is well under control, see left plot
of fig.~\ref{fig:bag}.
We refer to our contribution~\cite{Carrasco:2012dd} to the lattice 2012 conference
for additional details.
The heavy-quark mass dependence of $\omega_s$ is well described by the formula
$\omega_s(\overline\mu_h) = 1 + c_1(\lambda)/\overline\mu_h + c_2(\lambda)/\overline\mu_h^2$, 
see right plot of fig.~\ref{fig:bag}.
Applying the ratio-method one finally gets
\be
B_{B_s}^{\overline MS}(\overline{m}_b) = 0.90(5)\,,
\ee
where the error is the sum in quadrature of statistical and systematic uncertainties.
If we apply the ratio-method to $B_{B_d}$ and $B_{B_s}/B_{B_d}$ we obtain
the preliminary estimate of
\be
\frac{B_{B_s}}{B_{B_d}} = 1.03(2)\,, \qquad \qquad B_{B_d}^{\overline MS}(\overline{m}_b) = 0.87(5)\,.
\ee
Given this result we can also give a preliminary estimate for the parameter 
\be
\xi = \frac{f_{B_s}\sqrt{B_{B_s}}}{f_{B}\sqrt{B_{B}}}=1.21(6)\,,
\ee
where the error is the sum in quadrature of all uncertainties.

\section*{Acknowledgements}
\label{sec:ack}
\vspace{-0.4cm}
The computer time was made available to us by the Italian SuperComputing Resource Allocation (ISCRA) 
under the class A project HP10A7IBG7 "A New Approach to B-Physics on Current Lattices" and the 
class C project HP10CJTSNF "Lattice QCD Study of B-Physics" at the CINECA supercomputing service.
We also acknowledge computer time made available to us by HLRN in Berlin.

\bibliographystyle{h-elsevier}
\bibliography{bphys}

\end{document}